# Evaluating spintronics-compatible implementations of Ising machines

Andrea Grimaldi[1], Luciano Mazza[2], Eleonora Raimondo,[1] Pietro Tullo[2], Davi Rodrigues[2], Kerem Y. Camsari,[3] Vincenza Crupi[1], Mario Carpentieri[2], Vito Puliafito[2,*], Giovanni Finocchio[1,*]

[1] *Department of Mathematical and Computer Sciences, Physical Sciences and Earth Sciences, University of Messina, 98166, Messina, Italy*

[2] *Department of Electrical and Information Engineering, Politecnico di Bari, 70126, Bari, Italy*

[3]*Department of Electrical and Computer Engineering, University of California Santa Barbara, 93106, Santa Barbara, CA, USA*

**Abstract**

The commercial and industrial demand for the solution of hard combinatorial optimization problems push forward the development of efficient solvers. One of them is the Ising machine which can solve combinatorial problems mapped to Ising Hamiltonians. In, particular, spintronic hardware implementations of Ising machines can be very efficient in terms of area and performance, and are relatively low-cost considering the potential to create hybrid CMOS-spintronic technology. Here, we perform a comparison of coherent and probabilistic paradigms of Ising machines on several hard Max-Cut instances, analyzing their scalability and performance at software level. We show that probabilistic Ising machines outperform coherent Ising machines in terms of the number of iterations required to achieve the problem's solution. Nevertheless, high frequency spintronic oscillators with sub-nanosecond synchronization times could be very promising as ultrafast Ising machines. In addition, considering that a coherent Ising machine acts better for Max-Cut problems because of the absence of the linear term in the Ising Hamiltonian, we introduce a procedure to encode Max-3SAT to Max-Cut. We foresee potential synergic interplays between the two paradigms.

Corresponding authors: *gfinocchio@unime.it, *vito.puliafito@poliba.it

**I Introduction**

Physics-inspired unconventional computing paradigms [1–9] are receiving great attention in the scientific community in the search of hardware-friendly and scalable approaches for solving hard combinatorial optimization problems (COPs). COPs are usually characterized by a non-deterministic polynomial-time (NP) complexity [6]. Problems belonging to this class are usually solved with algorithms whose time-to-solution (TTS) scales exponentially with the size of the instances. Unconventional nondeterministic algorithms do not yet offer improvements to the performance of state-of-the-art conventional approaches. To overcome the current limits, a promising unconventional approach is based on the concept of Ising machines (IMs), devices that search the ground state of an Ising Hamiltonian, a problem that itself belongs to the NP-class for certain interaction matrix classes. It is possible to map COPs into Ising Hamiltonian in polynomial time [10]. Although IMs can be implemented in software, their high hardware-friendliness suggests that a specialized hardware compatible with the algorithm used to design the IM could lead to higher speed and efficiency. Two promising hardware implementations of IMs are coherent Ising machines (cIMs) [1,2] and probabilistic Ising machines (pIMs) [4,9,11], both of which are also spintronics-compatible as discussed below. Here, we show a comparison of the performance of those two paradigms on a representative example of NP-hard COPs that can be easily mapped to Ising Hamiltonians. We consider the calculation of the maximum cut (Max-Cut) of a graph [7]. Max-Cut consists in finding the bipartition of an undirected graph characterized by the largest sum of the weights of the edges between the two partitions. The solution process of a Max-Cut problem consists in the following steps: i) assigning a binary value to all the nodes; ii) evaluating the sum of the weights of all edges with different endpoints values, as shown in Fig. 1. This problem can be naturally encoded in an Ising Hamiltonian, corresponding to a quadratic Hamiltonian with discretized variables, similarly to a spin glass problem [8]:

$$H(\mathbf{s}) = -\sum_{ij} J_{ij} s_i s_j \ ; \tag{1}$$

where $\mathbf{s} = \{s_1, ..., s_i, ..., s_N\}$ is the vector of binary spin states of the system ($s_i \in \{-1, +1\}$) and $J$ is the symmetric coupling matrix of the graph. The matrix element $J_{ij}$ stores the weight of the edge joining nodes $i$ and $j$; the left part of Fig. 1 shows an example of construction of a $J$ matrix. In cIMs, the binary spin states are mapped to the relative phase of oscillators in the sub-harmonic injection locking regime. In pIMs, the binary spin states are implemented with a stochastic bistable system.

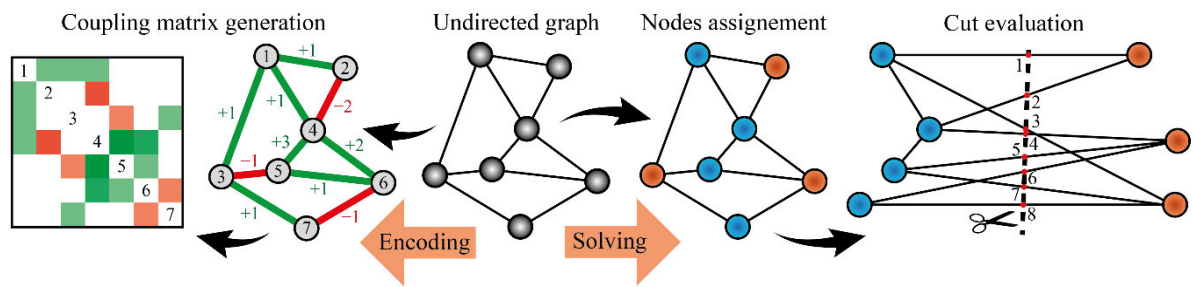

FIG. 1. A visual example of a maximum cut problem. The graph is encoded into an Ising Hamiltonian with a symmetric matrix $J$ whose values correspond to the weights of the connections. In an optimal solution, the nodes are in one of the two states in such a way that the bipartition "cuts" the highest value of accumulated edges.

The Ising spin of both cIMs and pIMs can be implemented with a CMOS-compatible magnetic tunnel junction (MTJ), with the final goal of having IMs that can be manufactured with a heterogenous or monolithic integration of hybrid spintronic/CMOS technology. In this respect, three-terminal MTJs are compatible with both cIMs and pIMs. The two IMs are then compared in terms of performance, highlighting advantages and disadvantages. Moreover, we envision their compatibility for potential synergic interplays between the two paradigms. We show that pIMs outperform cIMs in software implementations. Nevertheless, cIMs show a great potential for the high working frequency at which spintronic oscillators can operate [12].

The paper is organized as follows. Section II presents potential spintronic implementations of the Ising spin for cIMs and pIMs. Section III describes the main model used for the cIMs, i.e., the Kuramoto model, and compare it to the universal model of nonlinear oscillators with negative damping used to model spintronic oscillators [13] (here referred to as Slavin model). Section IV briefly introduces the concept of probabilistic computing with p-bits [9]. Section V compares cIMs with pIMs on a well-known benchmark instance library. Section VI tests one of probabilistic computing's encoding techniques, called invertible logic gates, on a modified Kuramoto model. Section VII evaluates the performance of a problem mapping that allows to take advantage of the cIM's good performance with Max-Cut to solve other hard COPs. Section VIII summarizes results and outlooks.

**II. Implementation of Ising spins for cIMs and pIMs with three-terminal magnetic tunnel junctions**

A promising platform for the implementation of Ising machines with low power consumption is the spintronic technology. To this end, here we show a versatile design based on three terminal MTJs [14], combining current-induced spin-orbit-torque (SOT) [15–17] and spin-transfer-torque (STT) [18–21], and voltage-controlled magnetic anisotropy (VCMA) [22,23], for the implementations of Ising spins. These effects, when properly combined, allow to successfully achieve binarization and tunability, enabling the implementation of several IM paradigms. This part of the work is based on micromagnetic simulations performed by numerically integrating the Landau-Lifshitz-Gilbert-Slonczewski equation [24]:

$$\frac{d\boldsymbol{m}}{dt} = \frac{\gamma M_S}{(1+\alpha_G^2)} \left( -(\boldsymbol{m} \times \boldsymbol{h}_{eff}) - \alpha_G (\boldsymbol{m} \times \boldsymbol{m} \times \boldsymbol{h}_{eff}) \right. \\ \left. + \sigma \big( J_{STT}\, g_T\, (\boldsymbol{m} \times \boldsymbol{m} \times \boldsymbol{p}) + J_{SOT} \theta_{SHE}\, (\boldsymbol{m} \times \boldsymbol{m} \times \boldsymbol{\sigma}_{SHE}) \big) \right) \quad (2)$$

where $\boldsymbol{m} = \frac{\boldsymbol{M}}{M_S}$ is the normalized magnetization vector, $\gamma$ is the gyromagnetic ratio, $M_S$ is the saturation magnetization of the MTJ free layer (FL), and $\alpha_G$ is the Gilbert damping factor. $\boldsymbol{h}_{eff}$ is the effective magnetic field, which includes the demagnetizing field and the interfacial uniaxial perpendicular anisotropy that also contains the VCMA contribution. The spin torques are proportional to a pre-factor $\sigma = \frac{g|\mu_B|}{|e|\gamma M_S^2 d_z}$, where $g$ is the gyromagnetic splitting factor, $\mu_B$ is the Bohr magneton, $e$ is the electron charge, and $d_z$ is the thickness of the FL. The STT is proportional to the spin polarization function $g_T$ [18], which is a function of the spin polarization $P$, and depends on the polarizer orientation, $\boldsymbol{p}$. The SOT is proportional to the spin hall angle $\theta_{SHE}$, a parameter that depends on the heavy metal (HM), and its orientation is determined by the unit vector $\boldsymbol{\sigma}_{SHE}$. The latter, for the Cartesian coordinate system fixed here, is along the $x$-direction, considering the charge current in the HM flowing along the $-y$-direction. Both $J_{STT}$ and $J_{SOT}$ are current densities. See Fig. 2 for more details.

For cIMs, the MTJ is designed to have a free layer with perpendicular magnetization and an in-plane polarizer (Fig. 2(a)). The device is biased with a large enough SOT in order to excite a self-oscillation state [14,25,26]. We consider an MTJ with an elliptical cross-section. The simulation parameters are summarized in Tab. I. In order to have an Ising spin, it is necessary to binarize the oscillator's phase. This can be achieved with VCMA-driven parametric excitations, as already proposed for spintronic oscillators with nano-constrictions and MTJ with two terminals [27,28]. Our calculations predict the robustness of the binarization technique and that it can be achieved in a range of frequencies close to two times the self-oscillation frequency. An example of parametric locking is shown in Fig. 2b. Here we can observe the binarization of the locked states (red and blue curves). The phase difference between the self-oscillation and the ac VCMA input (Fig. 2(b)) of the two possible synchronized states is shifted by 180 degrees. The binarization can also be achieved using an ac STT current that drives the injection-locking at the second harmonics (Fig. 2(c)) as already observed experimentally [27,29].

The implementation of pIMs requires the realization of a p-bit, i.e., a tunable stochastic bistable system [5,9]. These p-bits can be naturally implemented with stochastic MTJs, with the search of p-bits implementations with sub-nanosecond switching time being a very active research direction [30,31]. The first design of tunable p-bit based on a 3-terminal MTJ [9] has been demonstrated experimentally recently [32]. Here, we propose a p-bit implementation with three terminal MTJs based on the idea of magnetic clocking, as introduced for nanomagnetic logic [33,34] and physical unclonable functions [35]. The sketch of the MTJ for the p-bit implementation is described in Fig. 2(d). The optimized MTJ has a circular cross-section with the equilibrium magnetic state of both free and polarizer layers being out-of-plane and with no need of the VCMA effect. The tunable random number generation (TRNG) process is summarized in Fig. 2e with a representation of how the energy landscape of the FL magnetization changes in presence/absence of SOT current. The process goes as follows: i) before applying the SOT, the FL energy landscape has two stable minima along the $z$-axis, and the FL magnetization is in one of the two stable configurations. ii) After applying a large enough SOT, the energy landscape changes and it has a minimum in which the magnetization of the FL is aligned along the spin-current direction $y$-axis [35] iii) Once the SOT is switched off, the magnetization relaxes towards one of the two $z$-axis directions due to the perpendicular magnetic anisotropy, both of which are stable states, sampled with an equal probability. The STT from the third terminal can adjust this switching probability, finally resulting in a tunable p-bit, as shown in Fig. 2(f). Each point of Fig. 2(f) is obtained by averaging over $10^4$ realizations with different seeds. The points are well-described by a sigmoidal hyperbolic tangent.

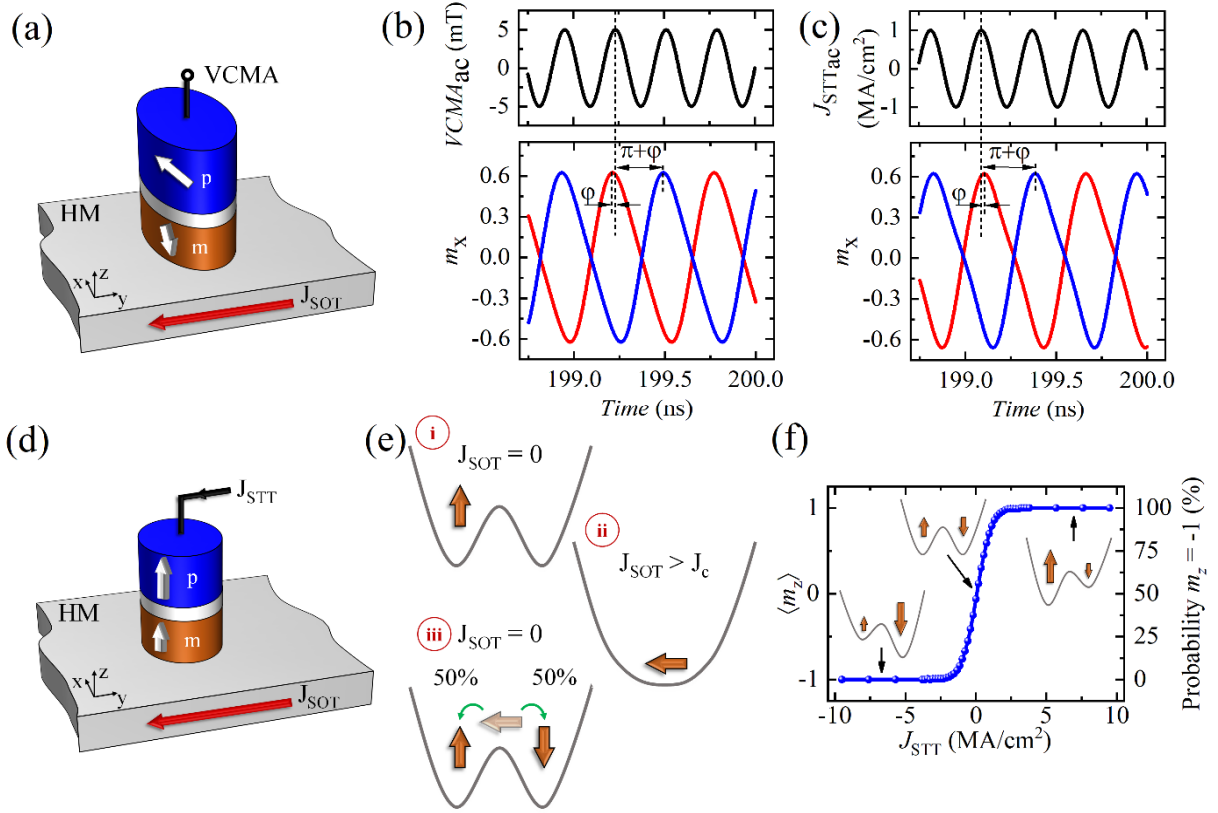

FIG. 2. (a) A schematic of an MTJ designed to behave as an oscillator for a cIM. The MTJ has three terminals and has an elliptical cross section with the polarizer having its magnetization aligned along the $x$-direction. The self-oscillations are driven by the dc current density and the VCMA signal acts as the injection-locking signal. (b) In the top panel, the VCMA signal acting on the MTJ; in the bottom panel, $m_x$ dynamics from simulations of two distinct devices, showcasing the phase binarization phenomenon due to the injection-locking from the VCMA. (c) The same behavior shown in (b) can be achieved by using an ac STT current density as the injection-locking signal. (d) A schematic of an MTJ designed to behave as a p-bit for a pIM. The MTJ has three terminals and is a circular cylinder with an out-of-plane polarizer. (e) Schematic showing the three-steps process of tunable random number generation (RNG) as described in the text. The dependence of this probability on the signal intensity is sigmoidal. Each point of this plot is the average of $10^4$ RNG simulations. The energy landscape of the three main states is shown in the insets.

TABLE I. Parameters used in the LLG micromagnetic simulations.

| LLG micromagnetic model | | |
| --- | --- | --- |
|  | cIM | pIM |
| $Size$ ($nm^3$) | $100 \times 40 \times 1$ | $50 \times 50 \times 1.4$ |
| $M_s$ ($MA/m$) | 1.10 | 0.80 |
| $K_u$ ($MJ/m^3$) | 0.62 | 0.060 |
| $J_{STT_{dc}}$ ($MA/cm^2$) | – | $[-10, 10]$ |
| $J_{STT_{ac}}$ ($MA/cm^2$) | $-1.00$ | – |
| $J_{SOT}$ ($MA/cm^2$) | $-1.51$ | $100$ |
| $VCMA_{ac}$ ($mT$) | $5.00$ | – |
| $f_0$ ($GHz$) | $1.64$ | – |

**III. Modeling cIMs**

State-of-the-art implementations of cIMs exploit coupled oscillators realized with various physical systems [1,36–38], and most of them employ the Ising Hamiltonian $J$ matrix as the coupling matrix of the system, achieving the binarization of the phase by sub-harmonic parametric locking at 0 an $\pi$ radians, as already discussed in previous sections. Hence, a given Max-Cut instance is solved using a network of oscillators that naturally evolves toward a low energy state corresponding to the problem's solution, set by the coupling matrix. Optical parametric oscillators obtain solutions very quickly [1,36], however the size of the required systems is not suitable for highly integrated applications. LC oscillators have been successfully used to solve Max-Cut problems [39,40], with the advantage being the possibility of performing rapid analysis with lumped components, but the use of inductors and capacitors also hinder highly integrated solutions. Similar obstacles have been shown for insulator to metal (IMT) phase-transition nano-oscillators (PTNO) which provide excellent energetic performances, but require capacitors to couple devices [41,42].

Spintronic solutions have been simulated [43,44] and recently an 8-spin Max-Cut solver was realized in hardware using spin hall nano-oscillators (SHNO). The main limitation observed was overcoming the slow propagation of spin waves [45], although scalability is still an issue. Considering the state-of-the-art and future perspectives, MTJ-based oscillators offer an optimal solution for forthcoming implementations of an Ising Machine chip for daily life applications due to their exploitable nonlinear dynamics, low sizes, GHz working frequencies and CMOS compatibility.

In the following, we describe two oscillator models and test their performance with the goal to furnish the tools and directions to design a scalable and compact cIM.

**A) Kuramoto model**

The Kuramoto model was first introduced to provide a general description of synchronization [46,47]. By adding a binarizing term and designing the coupling matrix, the generalized Kuramoto model provides a simple analytical description of cIMs. In this generalized model, the phases of the coupled oscillators interdependently evolve according to $N$ coupled differential equations, each of them describing the dynamics of the phase of a single oscillator $\phi_i$. For the i$^{\text{th}}$ oscillator we have:

$$\frac{d}{dt}\phi_i(t) = -K \sum_{j=1, j\neq i}^{N} J_{ij}\sin\left[\phi_i(t) - \phi_j(t)\right] - S\sin[2\phi_i(t)] + \xi(t); \qquad (3)$$

where $\boldsymbol{\phi} = \{\phi_i, \ldots, \phi_N\}$ represents the phases of the coupled oscillators, $K$ is the coupling strength, $S$ is the parametric locking parameter and $\xi$ an additive white Gaussian phase noise [48]. We solved the dimensionless dynamical equations, as it does not influence the accuracy of the Ising machine. As already discussed, the parametric locking term applies a signal with twice the frequency of the oscillator to binarize the oscillator's phase (sub-harmonic locking) into multiples of $\pi$ radians.

The advantage of this model lies in its easy-to-integrate dynamical equations that allows the implementation of advanced annealing schemes with relatively minor effort. Fig. 3(a) shows the solution of an exemplary 100 oscillator cubic Max-Cut problem using a linear annealing scheme of the coupling strength $K$ and the parametric locking $S$. Both coefficients are multiplied by an annealing parameter $C$ that starts at zero, and linearly grows up to one. For each run, this annealing scheme is applied five times, consecutively. The annealing scheme and an example trend of the Max-Cut as a function of time are shown in Fig. 3(b). $K$ and $S$ are selected heuristically in order to optimize the performance of the simulated cIM and are, in general, dependent on the topology of the problem. Each of the oscillators' phases are initialized with a random value between 0 and $\pi$ radians. Fig. 3(a) also shows how the system quickly reaches a local minimum, remains stable as long as the annealing parameters are below a well-defined threshold, which depends on a trade-off between the coupling strength and the force driving the phase binarization. This annealing strategy allows an effective exploration of the energetic landscape, with every reset of the coupling coefficients often resulting in a better Max-Cut.

The Kuramoto model has already been tested against large-scale benchmark problems and has shown very promising results [48]. It can be used to describe spintronic oscillators with reduced frequency-power coupling [49,50] or characterized by soliton dynamics, such as vortices [51,52] and bubbles [53]. However, this simple model cannot fully describe the physical behavior of oscillators in which phase and power are coupled. For example, this is the case of spintronic oscillators where, as a first approximation, phase and power are linked through the nonlinear frequency shift $N_0$ [13,49,50,54], as described in detail in the following section.

**B) Universal model of nonlinear oscillators with negative damping (Slavin model)**

Experiments have shown the accuracy of the Slavin model to simulate and understand the behaviour of spin-torque nano oscillators and spin-hall oscillators where the oscillator power and phase are coupled [53,55], and their collective behaviors when they interact [43]. The dynamics of each oscillator can be described by two coupled differential equations [13],

$$\frac{dp_i}{dt} = -2p_i[\Gamma_{+,i}(p_i) - \Gamma_{-,i}(p_i)] + 2F_e\sqrt{p_i}\cos(2\omega_i t + 2\phi_i) +$$
$$+2\Omega \sum_{j,j\neq i}^{N} J_{ij}\sqrt{p_i p_j}\cos(\phi_i - \phi_j - \beta) + \xi_p(t), \tag{4a}$$

$$\frac{d\phi_i}{dt} = -\omega_i(p_i) - \frac{F_e}{\sqrt{p_i}}\sin(2\omega_i t + 2\phi_i) + \Omega \sum_{j,j\neq i}^{N} J_{ij}\sqrt{\frac{p_i}{p_j}}\sin(\phi_i - \phi_j + \beta) + \xi_\phi(t), \tag{4b}$$

where $\phi_i(t)$, and $p_i(t)$ describe the time evolution of the oscillator power and phase, respectively. In both equations, the first term on the righthand side describes the dynamics of an independent device. $\Gamma_+$ and $\Gamma_-$ are functions representing respectively the positive and negative damping effects. First order expansion of these functions results in $\Gamma_{+,i} = \Gamma_G(1 + Qp_i)$ and $\Gamma_{-,i} = \Gamma_G I_{\text{ratio}}(1 - p_i)$, where $Q$ is the nonlinear damping coefficient, and $I_{\text{ratio}}$ is the ratio between the applied current and the threshold current necessary to excite self-oscillations. These expressions are proved to accurately describe experimental findings [13,55,56]. The frequency $\omega_i$ of each oscillator is linked with its power $p_i$ through the relation $\omega_i = \omega_0 + N_0 p_i$, where $\omega_0$ is the resonance frequency and $N_0$ is the nonlinear frequency shift. The noise has been implemented as proposed in [13], and the results obtained with and without the application of the thermal field at room temperature are qualitatively equivalent.

The term with amplitude $F_e$ is the external signal used for the parametric locking. Finally, the third term models the effect of the coupling between the oscillators and is dependent on the coupling strength $\Omega$ and the topology of the oscillator network, with $J_{ij}$ being the corresponding element of the Ising matrix (see Eq. (1)). The $\beta$ parameter represents the phase delay of the two coupled signals, it depends mainly on the coupling mechanism and the spatial distance among oscillators when considered [57].

These models can be used for the implementation of spintronic cIMs [43]. The parameters used for the simulations are based on a experimental parameters of MTJ-based spintronic oscillators [58] having CoFeB as a free layer (see the whole set of parameters in Tab. II). Fig. 3(c) shows the solution of the same problem of Fig. 3(a) obtained solving Eq. (4a) and (4b). In this case the phases and the powers of the oscillators are shown, even if only the phase is used to evaluate the Max-Cut. We notice that, as opposed to the simulation of the Kuramoto model of Fig. 3(a), the exploration of the phase landscape (Fig. 3(c), upper panel) takes place during the first part of the annealing period, with the network dynamics stabilizing toward the end. The plot of the powers shows their relationship with the annealing schedule (Fig. 3(c), lower panel) and, as the phases, they tend to binarize. At the beginning of the new annealing cycle, the metastable state becomes unstable, and the powers drop rapidly. This simple parameter schedule improves the Max-Cut as the number of annealing cycles increases.

**C) Comparison between Kuramoto and Slavin models**

Fig. 3(b) shows the cut values as a function of time of the models over the same instance. Both Kuramoto and Slavin cIMs reach a maximum value of 136. Overall, the former gets to an energetic minimum more rapidly, while the latter has a wider exploration of the energetic landscape. In both cases, the annealing schedule (also shown in Fig. 3(b)) is a key component of the energy-minimization process.

A systematic comparison between the performance of software implementation of cIMs (Kuramoto vs Slavin models) has been performed considering Max-Cut instances with an increasing number of nodes (each node is an oscillator). The Max-Cut achieved for each instance is shown in Fig. 3(d) and reveals that the accuracy of the two models is comparable. In other words, the coupling between power and phase does not reduce the performance of a cIM. This is one of the main results of this work.

We have also performed a grid search of the optimal nonlinear frequency shift $N$ and nonlinear damping coefficient $Q$, running with all the models the same 100 random instances of Max-Cut problems of cubic graphs with 100 oscillators and averaging the results. We emphasize that the tunable parameter $N$ measures the coupling between the power and phase dynamics and is a key difference between the two studied models [13]. For $N$ equals to zero, the frequency of the oscillators in the Slavin model is independent of the power as in the Kuramoto model.

Tab. III shows that, in general, the parameter $Q$ does not influence the Max-Cut score as much as the parameter $N$, for which an optimal value is approximately $N = 10N_0$, where $N_0$ is a reference value [43]. These results underline that the nonlinearities of spintronic oscillators might be beneficial for the realization of an IM hardware implementation and reveal that, by properly tuning the system, the Slavin model can achieve better results than the Kuramoto model.

TABLE II. Parameters used in the simulations of the Slavin model Ising machine.

| **Slavin model** |
| --- |

| | |
|---|---|
| $\omega_0$ (GHz) | 4.2 |
| $N/2\pi$ (GHz) | $-3.44$ |
| $Q$ | 2 |
| $I_{ratio}$ | 2 |
| $\Gamma_G$ (MHz) | 252 |
| $\beta$ (rad) | $-0.64\pi$ |
| $F_e$ | $3 \times 10^9$ |
| $\Omega$ | $1 \times 10^9$ |
| $dt$ (ps) | 5 |

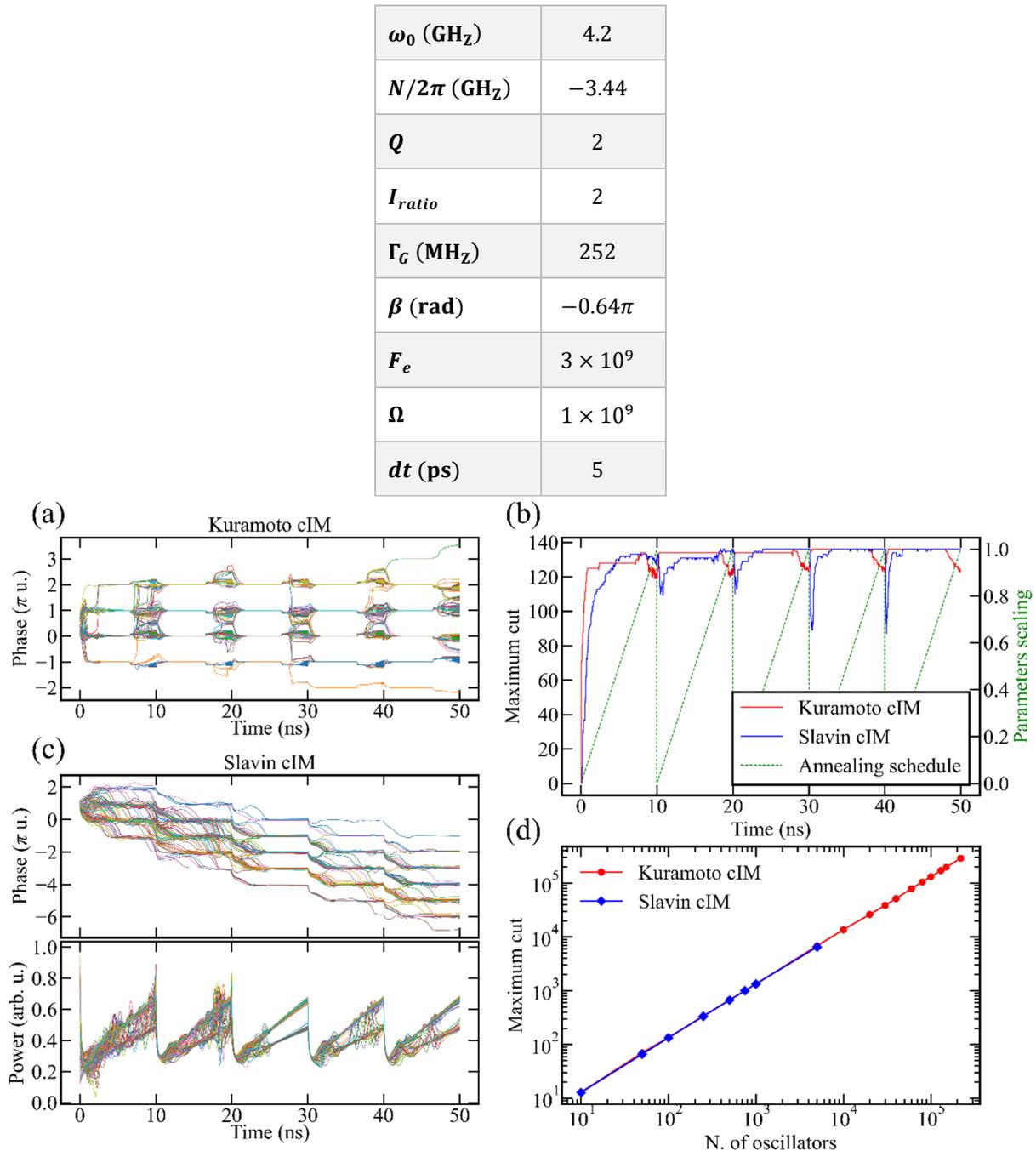

FIG. 3. (a) – (c) Example runs of maximum cut search of the same randomly generated cubic graph with 100 oscillators simulated using the Kuramoto model (a) and the Slavin model (c). The plots show the phases, used to evaluate the cut value, and the powers of the oscillators in the latter case. The cut values of both models evaluated in each time step (continuous line) are plotted in (b) together with the linear annealing schedule of the respective annealing parameters (dashed line). Both models use a saw-tooth annealing schedule. (d) A comparison between the performance of the Kuramoto model and the Slavin model finding the Max-Cut of randomly generated cubic graphs. For most of the reported values, the Max-Cut has been averaged over the solution of 100 different randomly generated graphs; for a higher number of oscillators only one run is considered.

TABLE III. Grid search of the optimal values of nonlinear frequency shift $N$ and nonlinear damping coefficient $Q$. The Max-Cut has been evaluated averaging the result obtained inputting the same 100 randomly generated cubic graphs with 100 oscillators to the model with the many combinations of values.

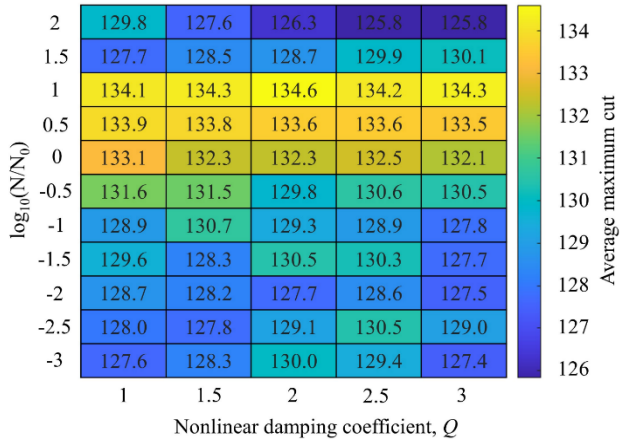

## IV. Modeling pIMs

In pIMs, the nodes of the Ising model are represented by p-bits which can be hardware-implemented with stochastic MTJs [9] or, as proposed in this work, with a deterministic clocked excitation of the randomness. The update process of the p-bits in a pIM is described by the following equations [5,9],

$$m_i(t) = \text{sgn}(\text{rand}(-1,+1) + \tanh(I_i(t))), \quad (5a)$$

where

$$I_i(t) = I_0(t)\left(\sum_{ij} J_{ij} m_j(t) + h_i\right). \quad (5b)$$

Here $\boldsymbol{m} = \{m_1, \ldots, m_N\}$ is the state vector of the probabilistic bits (p-bits), $I_0$ is a pseudo-temperature parameter used to control the annealing process and $\boldsymbol{h} = \{h_1, \ldots, h_N\}$ is the bias vector, an additional term meant to represent external field excitations in the general formulation of an Ising Hamiltonian. By updating a system's p-bits in sequence, the state samples the energy landscape of the Ising Hamiltonian. Using common annealing strategies like simulated annealing [59] or parallel tempering [60,61], the pIM solver drives the system to the ground state [5].

pIMs have been employed successfully with COP encodings using probabilistic spin logic (PSL) [4,9]. PSL's strategy is to use invertible logic gates, Ising models whose energy function has its minima corresponding to the truth table state of the gate. Their advantage is that, since they are transposed to an undirected graph, they no longer have a preferential input/output combination and can thus be operated in reverse. This allows for a conceptually simple way to solve COPs: by using a logic circuit where the standard logic gates are substituted by invertible logic gates, the solution can be reached by operating a simple circuit in reverse [9].

## V. A comparison between cIMs and pIMs

Simulated performance of cIMs based on the Kuramoto model (similar results are also obtained with the Slavin model) and pIMs were compared by attempting to solve several hard Max-Cut instances from G-set graphs, generated by the machine independent graph generator "rudy" of G. Rinaldi. Each instance was attempted one-hundred times by both solvers and the cut value as a function of the iterations was recorded. Both solvers were operated using heuristically optimized parameters. Tab. IV summarizes the results of the comparison. For each instance, we report both the average Max-Cut score and the best one achieved among the one-hundred runs. In addition, we also show the number of times the best result was reached, along with how often a solution close to the best one was obtained [48]. While there is variation between individual runs, the plot clearly shows that the software implementation of pIMs perform consistently better than cIMs. Fig. 4 shows the analysis of a single instance. The pIM runs (in red) reach higher-quality states than cIM ones (in blue) and in substantially shorter times.

However, it is important to note that we make the comparisons in terms of number of computational steps performed. In this context, pIMs, with their discrete dynamics, explore new states at a faster rate than cIMs, whose dynamics is obtained as a numerical integration of continuous differential equations. In a potential hardware implementation, the performance of the two paradigms depends largely on the specific properties of the devices used for the implementation. In this context, we envision that the two paradigms can achieve comparable performance, and even effectively synergize if used in combination.

TABLE IV. Results of solving attempts of instances of the G-set performed with both pIMs and cIMs. 'pIM (cIM) Best' is the best solution found among the 100 trials of each instance, '# of best' is the number of times that best result was achieved, and '# 0.999% of best' is the number of times a state with maximum cut 99.9 % close to the best result was reached.

| Instance name | pIM Average | | pIM Best | pIM # of best | pIM # 0.999% of best | cIM Average | | cIM Best | cIM # of best | cIM # 0.999% of best |
|---|---|---|---|---|---|---|---|---|---|---|
| g1 | 11571.06 | ± 19.30 | 11624 | 2 | 4 | 11475.38 | ± 25.01 | 11540 | 1 | 1 |
| g2 | 11571.90 | ± 17.28 | 11620 | 2 | 3 | 11481.84 | ± 31.89 | 11575 | 1 | 2 |
| g3 | 11573.58 | ± 19.20 | 11617 | 1 | 4 | 11473.73 | ± 29.95 | 11550 | 1 | 1 |
| g4 | 11601.52 | ± 23.56 | 11646 | 2 | 10 | 11497.78 | ± 37.83 | 11605 | 1 | 1 |
| g5 | 11585.88 | ± 18.77 | 11622 | 2 | 11 | 11491.06 | ± 30.18 | 11545 | 1 | 8 |
| g6 | 2128.77 | ± 20.56 | 2175 | 1 | 4 | 2029.92 | ± 29.10 | 2113 | 1 | 1 |
| g7 | 1965.63 | ± 18.10 | 2000 | 1 | 1 | 1866.66 | ± 31.79 | 1923 | 3 | 4 |
| g8 | 1967.06 | ± 16.84 | 1998 | 1 | 2 | 1878.82 | ± 29.02 | 1934 | 1 | 1 |
| g9 | 2007.54 | ± 18.50 | 2047 | 1 | 1 | 1908.26 | ± 37.49 | 1976 | 1 | 1 |
| g10 | 1958.16 | ± 17.60 | 1999 | 1 | 1 | 1864.34 | ± 33.02 | 1946 | 1 | 1 |
| g11 | 556.04 | ± 2.83 | 562 | 3 | 3 | 483.76 | ± 7.67 | 502 | 1 | 1 |
| g12 | 547.42 | ± 3.09 | 554 | 4 | 4 | 478.90 | ± 8.36 | 492 | 6 | 6 |
| g13 | 572.14 | ± 2.72 | 580 | 1 | 1 | 503.06 | ± 9.51 | 522 | 1 | 1 |
| g14 | 3045.17 | ± 4.43 | 3053 | 2 | 19 | 2967.60 | ± 10.39 | 2991 | 1 | 2 |
| g15 | 3029.81 | ± 6.25 | 3049 | 1 | 2 | 2947.04 | ± 9.63 | 2967 | 1 | 3 |
| g16 | 3031.10 | ± 5.15 | 3043 | 1 | 5 | 2953.35 | ± 10.34 | 2976 | 1 | 3 |
| g17 | 3026.31 | ± 5.02 | 3042 | 1 | 1 | 2948.26 | ± 10.73 | 2967 | 3 | 6 |
| g18 | 971.01 | ± 9.43 | 988 | 1 | 1 | 902.29 | ± 18.50 | 941 | 1 | 1 |
| g19 | 885.29 | ± 9.57 | 904 | 1 | 1 | 815.12 | ± 18.30 | 856 | 1 | 1 |
| g20 | 919.75 | ± 10.93 | 940 | 1 | 1 | 847.73 | ± 17.25 | 907 | 1 | 1 |
| g21 | 906.61 | ± 9.89 | 928 | 1 | 1 | 838.35 | ± 19.32 | 881 | 1 | 1 |
| g22 | 13278.04 | ± 21.72 | 13336 | 1 | 2 | 13060.89 | ± 35.50 | 13141 | 1 | 4 |
| g23 | 13282.67 | ± 18.81 | 13333 | 1 | 3 | 13071.73 | ± 34.54 | 13149 | 1 | 3 |
| g24 | 13271.28 | ± 18.02 | 13317 | 1 | 5 | 13063.72 | ± 32.00 | 13127 | 1 | 5 |
| g25 | 13273.72 | ± 20.04 | 13318 | 1 | 9 | 13075.01 | ± 34.39 | 13164 | 1 | 2 |
| g26 | 13263.88 | ± 18.27 | 13314 | 1 | 3 | 13063.50 | ± 32.36 | 13130 | 1 | 4 |
| g27 | 3272.67 | ± 19.28 | 3308 | 1 | 4 | 3066.10 | ± 38.15 | 3153 | 1 | 2 |
| g28 | 3234.82 | ± 17.94 | 3272 | 2 | 3 | 3026.49 | ± 34.10 | 3146 | 1 | 1 |
| g29 | 3334.90 | ± 14.52 | 3369 | 2 | 4 | 3130.31 | ± 33.79 | 3222 | 1 | 1 |
| g30 | 3348.80 | ± 19.76 | 3392 | 1 | 1 | 3137.66 | ± 36.18 | 3217 | 1 | 1 |
| g31 | 3245.56 | ± 19.65 | 3280 | 1 | 4 | 3040.71 | ± 34.89 | 3120 | 1 | 1 |
| g32 | 1382.90 | ± 5.41 | 1394 | 2 | 2 | 1202.66 | ± 14.35 | 1242 | 1 | 1 |
| g33 | 1358.48 | ± 4.41 | 1368 | 4 | 4 | 1181.68 | ± 13.76 | 1216 | 1 | 1 |
| g34 | 1361.26 | ± 4.60 | 1372 | 1 | 1 | 1186.12 | ± 14.71 | 1218 | 1 | 1 |
| g35 | 7633.99 | ± 8.70 | 7665 | 1 | 2 | 7430.10 | ± 19.89 | 7475 | 1 | 1 |
| g36 | 7626.45 | ± 10.29 | 7648 | 1 | 8 | 7422.34 | ± 16.98 | 7460 | 1 | 3 |
| g37 | 7634.26 | ± 9.28 | 7658 | 1 | 5 | 7435.55 | ± 17.99 | 7493 | 1 | 1 |
| g38 | 7634.18 | ± 8.48 | 7655 | 1 | 5 | 7432.76 | ± 17.15 | 7470 | 1 | 5 |
| g39 | 2350.69 | ± 14.06 | 2386 | 1 | 1 | 2173.84 | ± 29.96 | 2240 | 1 | 1 |
| g40 | 2336.12 | ± 16.45 | 2377 | 1 | 1 | 2154.72 | ± 34.09 | 2232 | 1 | 1 |
| g41 | 2337.63 | ± 15.59 | 2371 | 1 | 1 | 2160.42 | ± 26.21 | 2224 | 2 | 2 |
| g42 | 2416.33 | ± 15.26 | 2446 | 1 | 1 | 2240.86 | ± 28.05 | 2310 | 1 | 1 |
| g43 | 6625.06 | ± 15.04 | 6660 | 1 | 2 | 6526.77 | ± 21.25 | 6580 | 1 | 2 |
| g44 | 6620.06 | ± 13.36 | 6646 | 1 | 11 | 6523.80 | ± 21.92 | 6602 | 1 | 1 |
| g45 | 6623.28 | ± 11.71 | 6652 | 1 | 2 | 6522.96 | ± 23.82 | 6569 | 1 | 4 |
| g46 | 6618.45 | ± 11.42 | 6645 | 1 | 3 | 6524.41 | ± 24.40 | 6572 | 1 | 3 |
| g47 | 6626.77 | ± 11.52 | 6655 | 1 | 2 | 6532.38 | ± 21.01 | 6598 | 1 | 1 |
| g48 | 5964.80 | ± 50.43 | 6000 | 67 | 67 | 5463.16 | ± 40.16 | 5558 | 1 | 1 |
| g49 | 5940.60 | ± 33.40 | 6000 | 16 | 16 | 5472.82 | ± 39.31 | 5560 | 1 | 1 |
| g50 | 5843.08 | ± 27.36 | 5880 | 1 | 6 | 5465.58 | ± 40.21 | 5582 | 1 | 1 |
| g51 | 3822.72 | ± 5.84 | 3836 | 1 | 6 | 3724.16 | ± 12.00 | 3753 | 1 | 2 |
| g52 | 3826.41 | ± 5.58 | 3836 | 3 | 17 | 3728.32 | ± 12.69 | 3770 | 1 | 1 |
| g53 | 3827.53 | ± 6.29 | 3843 | 1 | 3 | 3726.23 | ± 13.17 | 3760 | 1 | 2 |
| g54 | 3824.56 | ± 5.89 | 3844 | 1 | 1 | 3726.59 | ± 11.82 | 3763 | 1 | 1 |

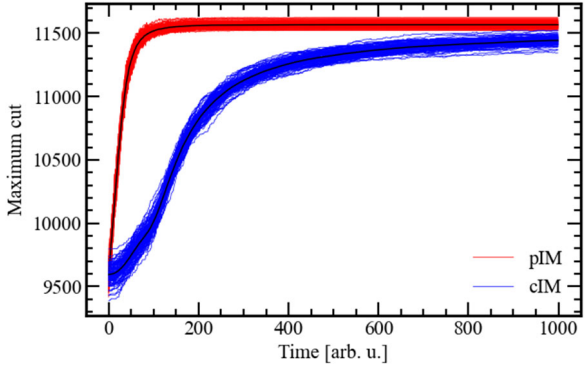

FIG. 4. A comparison of the pIM and the Kuramoto-based cIM model on the instance "g1.rud" from the G-set. The instance has 800 nodes and 19176 edges. The red (blue) area represents an ensemble of 100 solving attempts using pIM (cIM). For both models, the black line represents the average of all solving attempts. The annealing parameters were chosen after a systematic study to ensure fairness in the comparison.

## VI. Invertible logic gates with modified Kuramoto model

To implement invertible logic gates with cIM models, we consider an additional bias vector $h$ in the model of Eq. (3). The extra bias favors one of the two binarized phases for each oscillator making the probability to select a specific Ising spin state tunable, similarly to the corresponding term of Eq. (5a) for pIMs [9]. It corresponds to the full Ising Hamiltonian with local applied fields,

$$H(\boldsymbol{s}) = -\sum_{ij} J_{ij} s_i s_j + \sum_i h_i s_i. \qquad (6)$$

The generalized Kuramoto model to describe the Hamiltonian of Eq. 6 is given by

$$\frac{d}{dt}\phi_i(t) = -K \sum_{j=1, j\neq i}^{N} J_{ij} \sin[\phi_i(t) - \phi_j(t)] - K_h h_i \sin[-\phi_i(t)] - S \sin[2\phi_i(t)] + \xi(t), \qquad (7)$$

where $K_h$ represents the bias coupling strength which is proportional to a sinusoidal term at the same frequency of the oscillator frequency. The added term mimics the PSL pivot structure used in pIMs that locally bias the switching probability in a sigmoidal fashion and is independent of the states of other nodes.

To study the properties of the modified proposed model, we ran an extensive parametric study on an AND gate Ising encoding to estimate the optimal values of the three coupling parameters $K$, $K_h$ and

$S$ in the Kuramoto model (Eq. (7)). To assess whether the final state of each schedule belonged to the truth table of the AND gate, we ran a total of $10^4$ annealing schedules with randomized initial conditions and recorded their final state. After finding the optimal set of parameters for a balanced AND gate, we considered composite circuits of AND gates to assess whether the parametrization was scalable.

Fig. 5(a) shows that the probability of achieving a state compatible with the clamped output of the AND gate does match the expected values. This means that, when the output is equal to 1 (orange bars in the graph), 111 is the only visited state; on the other hand, when it is clamped to 0 (blue bars in the graph), the states 000, 100, and 010 are equally explored. In Fig. 5(b) the same parameters were used to test two AND gates connected acting as 3-inputs AND gate. While the selected states are mostly correct, the probability distribution of the final states does not match the energy distribution of the model, even accounting for statistical fluctuations.

In the time evolution of the phase space representation of Fig. 5(c), we observed that the final state of a run with more than one ground state depends almost exclusively on its initial phase configuration. This intuitively resembles an attraction/repulsion electromagnetic-like model, with the charges being the binary configurations of the phase space. The attraction/repulsion effect of a configuration depends on how energetically advantageous it is. The oscillators act as a test charge cast in a random position $\boldsymbol{P_0}$ of the phase space and subject to a force equal to:

$$\boldsymbol{F_T} = \sum_{n=1}^{2^N} \frac{C_n}{d(\boldsymbol{P_0}, \boldsymbol{S_n})} (\boldsymbol{P_0} - \boldsymbol{S_n}) \ , \tag{8}$$

Where $\boldsymbol{S_n}$ is the $N$-dimensional vector that represents the position in the phase space of the n[th] configuration and $C_n$ is its attraction/repulsion coefficient. $d(\boldsymbol{x}, \boldsymbol{y})$ is the $N$-dimensional Euclidean distance between two points. In Fig. 5(d) we compare the results of the cIM implementation with the effective model introduced. The similarities between the results show that the explorable energy landscape of the cIM implementation is indeed very sensitive to the initial phase of each oscillator, as described by the model of Eq. (8). This may be due to the annealing process for cIMs, which only allows classical trajectories in the phase space and produces a particle-like behaviour of the relative phases. Non-linearities of the phase dynamics, such as the ones described by the Slavin model, and modified annealing processes could overcome the observed limitation, improving the invertible logic gates implementation with cIMs.

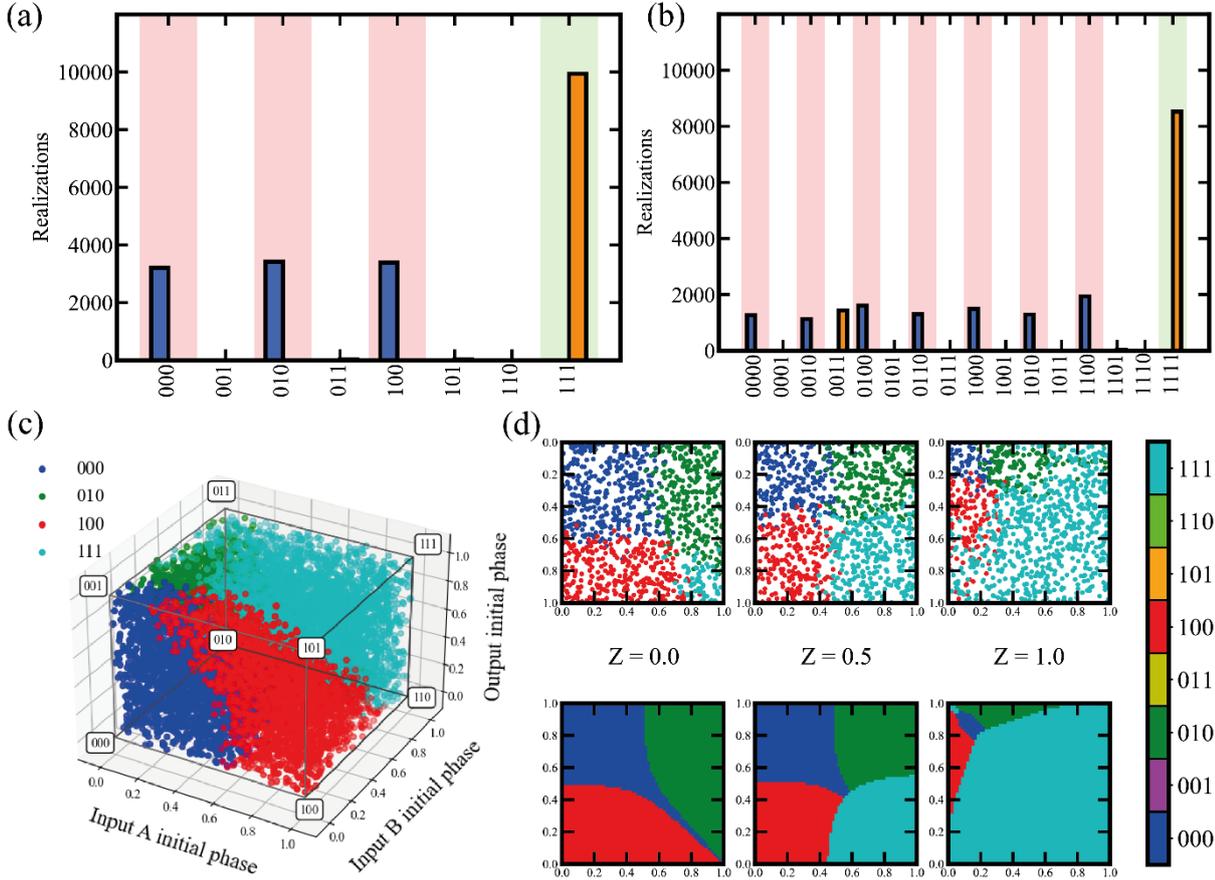

FIG. 5. (a)-(b) 2-inputs and 3-inputs AND gate simulation with optimized parameters $K = 1$, $K_h = 0.5$, $S = 1$. Each set of colored bars represents a simulation setup with a different bias $h$. In orange (blue) the output is clamped to $+1$ ($-1$), so $h_3$ is increased (decreased) by 10. The height of the bars is the number of times a randomized simulation has ended its annealing schedule in that state. The truth table states to visit with output clamped to $+1$ ($-1$) are highlighted in green (red). Composing circuits seems to affect the probability distribution, hindering scalability. (c) Three-dimensional representation of final state vs initial conditions of the unbiased AND gate simulated in (a). (d) Comparison between Kuramoto model oscillators and attraction model simulations. The similarities suggest that oscillators tend to gravitate toward the closest solution in the phase space.

**VII Maximum cut encoding of Max-3SAT**

The goal of the maximum satisfiability problem (Max-SAT) is to satisfy as many clauses as possible of an instance in conjunctive normal form. As mentioned earlier, COPs in the same complexity class can be mapped to each other. This allows, in principle, to solve Max-SAT problems by solving Max-Cut instances and vice versa. Due to the success of the cIMs to solve Max-Cut, we propose an alternative method to the use of invertible logic gates to solve Max-SAT problems by relying on the mapping between the two COPs. In particular, we demonstrate the solution of Max-3SAT (i.e., each clause has exactly three variables) by solving the equivalent Max-Cut. We emphasize that any Max-kSAT, with k >3, can be mapped to a Max-3SAT [62].

The mapping between Max-3SAT and Max-Cut for a single clause is graphically shown in Fig. 6(a). In the chosen map, we introduce additional ancillary nodes to mediate the interaction between the variable nodes to ensure that each clause with three variables is only satisfied if all three variables are true. By carefully choosing the couplings of the graph and by keeping one of the auxiliary oscillators fixed to phase zero, we ensure that a maximum cut solution corresponds to a satisfied clause. We mention that the inverse is not true, as some sets of variables values can satisfy a clause while not corresponding to the maximum cut of the graph. As illustrated by the example in Fig. 6(b), to join several clauses, one connects the shared variables nodes to the other clause nodes, inverting the sign of their couplings if the variable appears negated in that clause. Moreover, if two variables appear together in two different clauses, we consider the coupling as the sum of their couplings in each clause.

This mapping was used to obtain the graph of a simple Max-3SAT instance, "uf20-01.cnf", with 20 variables and 91 clauses. This instance was tested a total of 1000 times, each with a randomized initial configuration. The results are shown in Fig. 6(c) and 6(d). The system can achieve the optimal or close to it with high probability, although further testing is required to evaluate the scaling capabilities of the chosen map. In Fig. 6(d), we investigated how commonly the degeneracies caused by non-maximum cut states in satisfied clauses could bring to final guesses that reached the optimal solution while not achieving the maximum cut (few values with zero cost and maximum cut equal to 32). The opposite, achieving the maximum cut with a sub-optimal state, is indeed possible in a graph with several clauses, differently from the single clause case. As an example, if two clauses contain the same two variables, once with concordant and once with discordant signs, the coupling between the two cancels out and becomes zero, slightly altering the balance of those clauses' topology. However, as shown by the red line in Fig 6(d), the trend of the mapping shows that the degeneracy does not strongly influences the result of the Max-Cut, i.e., there is a strong correlation between the maximum cut and the Max-3SAT solution cost. The degeneracy of the map could be removed by considering sparse maps.

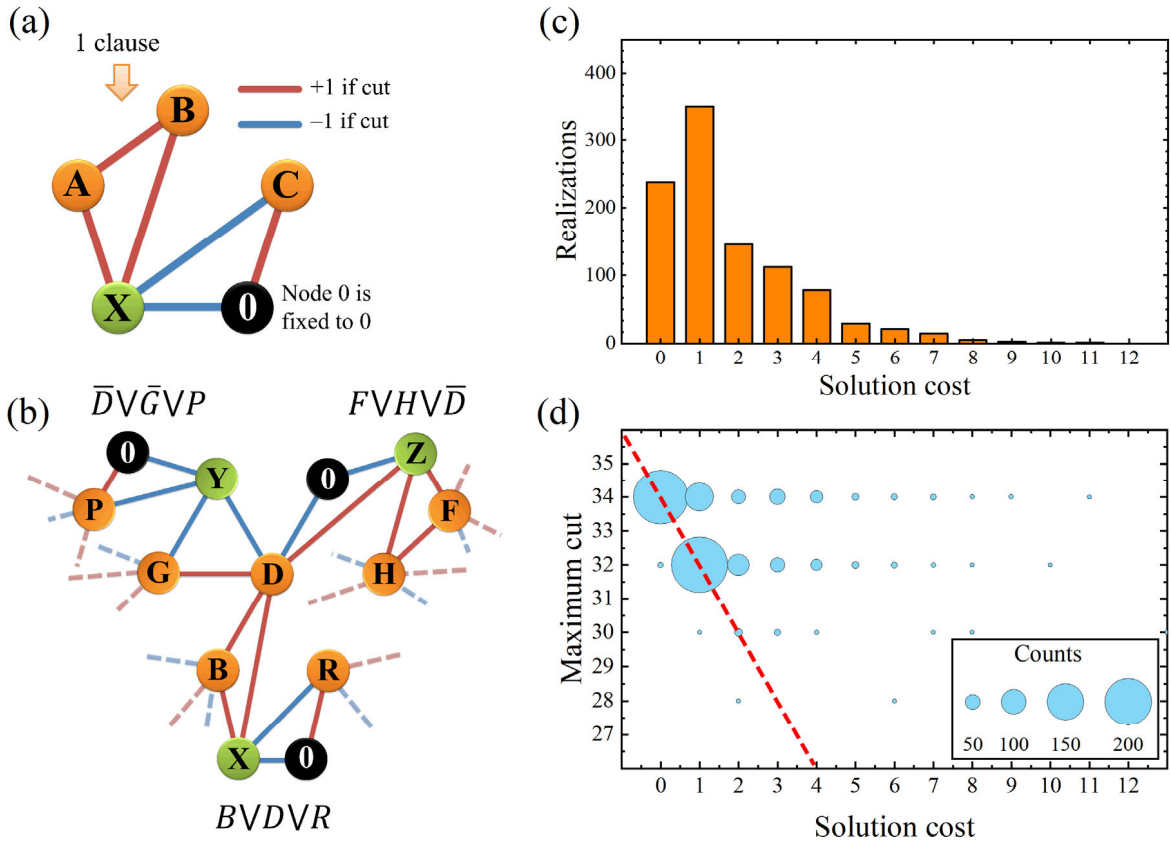

FIG. 6. (a) Graph representation of a single (A ∨ B ∨ C) logic clause. The maximum cut is only achieved in states where the clause is satisfied. (b) Graph of a toy Max-3SAT instance with an example of a variable shared in more than one logic clause. (c) – (d) States reached by the Kuramoto model system of oscillators for the simple Max-3SAT instance "uf20-01.cnf", with 20 variables and 91 clauses. We performed 1000 simulations with random initial state. The solution cost represents the number of logic clauses not satisfied at the last step of each simulation. The red dashed line is a guide to the eye to demonstrate the correlation between the maximum cut and the solution cost of the Max-SAT.

**Summary and Conclusions**

In this manuscript we have compare two different strategies which can be used for the hardware implementations of Ising machines, i) cIMs, based on a system of coupled oscillators, and ii) pIMs, based on a system of coupled p-bits, proposing a spintronic implementation of the Ising spin for each of them. From a modeling point of view, spintronic cIMs can be described by the Kuramoto model and the Slavin model. The latter includes the coupling between the oscillators' power and phase via the nonlinear shift. We compared the performance of the two models at software level and showed that the power-phase coupling may allow for a slightly better exploration of the cIM phase space, which can lead to higher accuracy. We also performed a comparison between the accuracy of pIMs and cIMs, showing that the first achieves better results. The use of invertible logic gates implemented with cIMs can underperform because of a classical-like particle behavior of the oscillators' phases. In addition, we proposed an alternative method to the use of invertible logic gates for the solution of Max-SAT, which exploits the effectiveness in solving Max-Cut instances of cIMs. The possibility to map between different COP problems in the same complexity class allows to leverage the optimal behavior of different cIM architectures and promises results for the implementation of cIMs to solve a wide range of COPs. Considering new optimized architectures and annealing schemes, which may include the combination of pIMs and cIMs, stimulates the development of fast, scalable, accurate and energy efficient hardware implementations of Ising machines.

**Acknowledgements**

This work was supported under the project number 101070287 — SWAN-on-chip — HORIZON-CL4-2021-DIGITAL-EMERGING-01, the project PRIN 2020LWPKH7 "The Italian factory of micromagnetic modeling and spintronics" funded by the Italian Ministry of University and Research (MUR), and by the PETASPIN association (www.petaspin.com). DR thanks the support from the project D.M. 10/08/2021 n. 1062 (PON Ricerca e Innovazione) funded by the Italian MUR. KYC acknowledges the support from a CNR YIP grant.